%% file: main.tex
\documentclass{IEEEtran}
\ifCLASSINFOpdf
   \usepackage[pdftex]{graphicx}
\else
   \usepackage[dvips]{graphicx}
\fi
\usepackage[cmex10]{amsmath}

\usepackage{graphics} % for pdf, bitmapped graphics files
\usepackage{graphicx}
\usepackage{subcaption}
\usepackage{array}
\usepackage{cite}
\usepackage{epsfig} % for postscript graphics files
\usepackage{amsmath} % assumes amsmath package installed
\usepackage{amssymb}  % assumes amsmath package installed
\usepackage{amsthm}
\usepackage{color}
\usepackage{float}
\usepackage[noabbrev]{cleveref}
\usepackage{algorithmic}
\usepackage{url}
\usepackage{calrsfs}
\usepackage{textcomp}
\usepackage{multirow}
\usepackage{booktabs}
\usepackage{bm} 

\DeclareMathAlphabet{\pazocal}{OMS}{zplm}{m}{n}
\DeclareMathAlphabet{\mathpzc}{OT1}{pzc}{m}{it}
\newcolumntype{P}[1]{>{\centering\arraybackslash}p{#1}}

\newcommand{\norm}[1]{\left\lVert#1\right\rVert}

\newcommand\ChangeRT[1]{\noalign{\hrule height #1}}

\usepackage[linesnumbered,ruled,vlined]{algorithm2e}
\SetKwInput{KwInput}{Input}                % Set the Input
\SetKwInput{KwOutput}{Output}              % set the Output
\SetKwInput{KwData}{Parameters}              % set the Output
\usepackage[cal = pxtx, scr = dutchcal]{mathalpha}

\begin{document}
\title{Understanding the Safety Requirements for Learning-based Power Systems Operations}

\author{Yize Chen, Daniel Arnold, Yuanyuan Shi and Sean Peisert

	\thanks{Y. Chen, D. Arnold and S. Peisert are with the Lawrence Berkeley National Laboratory,  emails: \{yizechen, dbarnold, sppeisert\}@lbl.gov. Y. Shi is with the University of California San Diego, email: yyshi@eng.ucsd.edu. This work was supported by Laboratory Directed Research and Development (LDRD) funds provided by Lawrence Berkeley National Laboratory, operated for the U.S. Department of Energy under Contract No. DE-AC02-05CH11231.}
}

\maketitle

\begin{abstract}
\input{abstract}
\end{abstract}


\section{Introduction}
\label{sec:intro}
\input{intro}

\section{Power System Operation Models}
\label{sec:system}
\input{system}

\section{Designing Learning-Based Controllers}
\label{sec:RL}
\input{rl.tex}

\section{Threat Model}
\label{sec:simulation}
\input{attack.tex}

\section{Case Study}
\label{sec:simulation}

\input{result.tex}

\section{Conclusion and Discussion}

\input{conclusion.tex}

\bibliographystyle{IEEEtran}
\bibliography{references}

\end{document}

%% file: abstract.tex
Recent advancements in machine learning and reinforcement learning have brought increased attention to their applicability in a range of decision-making tasks in the operations of power systems, such as short-term emergency control, Volt/VAr control, long-term residential demand response and battery energy management. Despite the promises of providing strong representation of complex system dynamics and fast, efficient learned operation strategies, the safety requirements of such learning paradigms are less discussed. This paper explores the design requirements on both data and model side of such learning algorithms by exploiting the impacts of adversarial attacks on safety critical system operations. Case studies performed on both voltage regulation and topology control tasks demonstrated the potential vulnerabilities of the standard reinforcement learning algorithms, and possible measures of machine learning robustness and security are discussed for power systems operation tasks.

%% file: intro.tex
The increasing penetration of renewables and the evolution of demand-side power load profiles have called for faster and more efficient decision-making for the complex power grids.  A fundamental challenge is that system operators need to find solutions for large-scale operational problems with hard time constraints, while such decisions should ensure the reliable supply of electricity and satisfy the economic as well as control objectives under various timescales. Machine learning (ML) algorithms such as end-to-end reinforcement learning (RL) have emerged as potential candidates, because of their capabilities on representing rich system states and fast inference procedure for finding the operation decisions directly from massive amount of data~\cite{marot2020learning}. Compared to standard model-based control and optimization pipelines, these learning-based algorithms can achieve higher precision in modeling complex power system dynamics, better accommodation of renewable generation, faster and more accurate response to uncertainty and variability~\cite{ernst2004power, dobbe2020learning}.

\begin{figure}[t]
	\centering
	\includegraphics[width=0.99\linewidth]{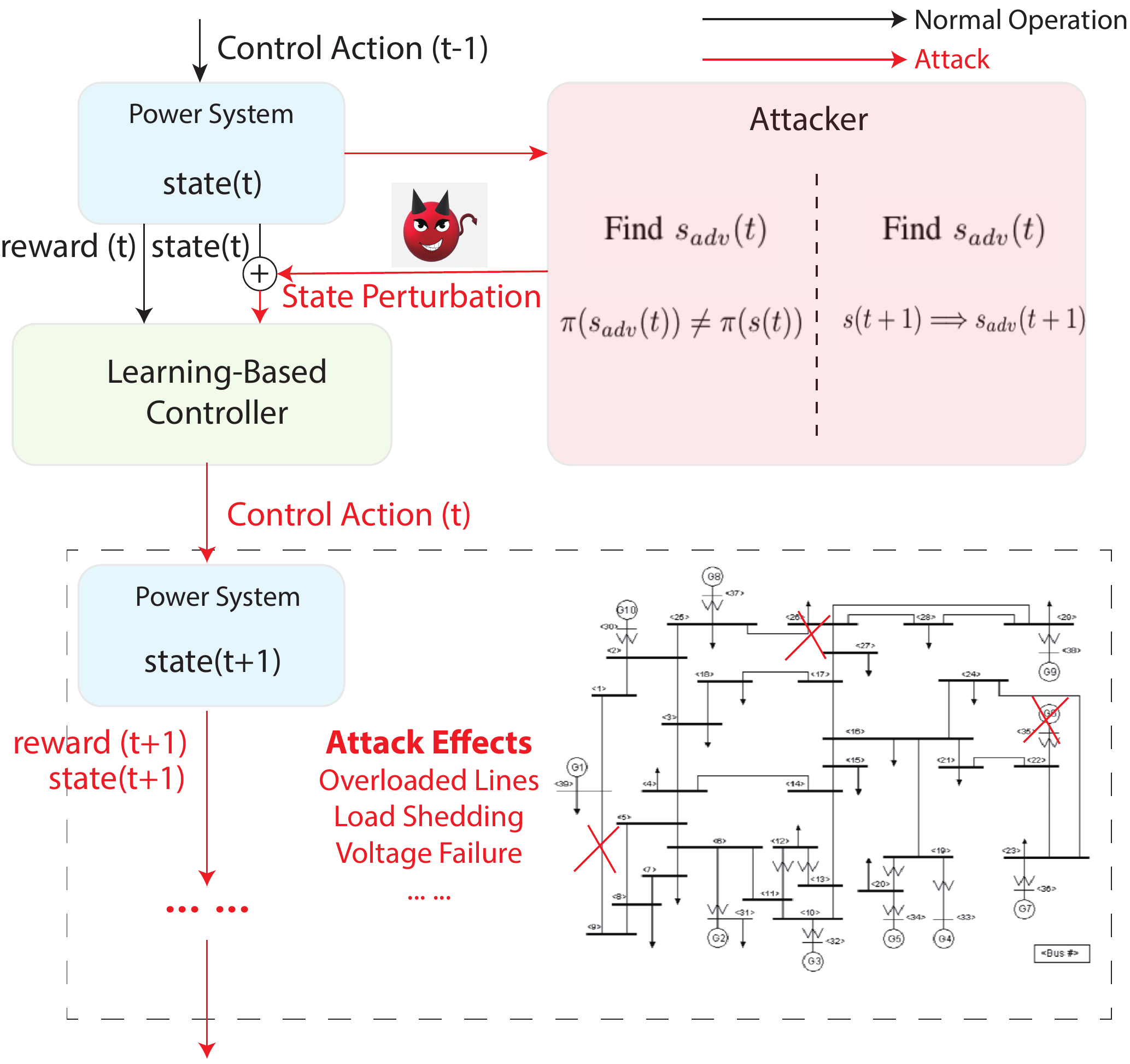}
	\caption{\footnotesize Illustration of our proposed attack strategy on learning-based controllers. Compared to normal training and operating scheme for a data-driven power system controller, we evaluate the black-box attack via crafting two kinds of adversarial input data perturbation for the learning agents. Such overlooked attacks can bring a number of threats to the grid operation.}
	\label{fig:workflow}
\end{figure}

However, researchers have noticed that deep and complex models achieving high performance on clean training data are often vulnerable to small adversarial perturbations over input space~\cite{goodfellow2014explaining, huang2017adversarial}. Specifically designed, unnoticeable noises added on measurements can largely modify the ML model output. As power system operations are safety critical, system failures caused by incorporating data-driven processes are not favorable. More importantly, due to the sequential nature of power system dynamics, adversarially perturbed decisions can cause a series of damage such as overloaded power lines and cascading failure. This is a real concern.  Yet few research have been discussing the cyber and data threats brought by ML automation~\cite{chen2018machine}.
This motivates us to raise the question: 

% \emph{Are power system operation schemes learned directly from data reliable?}
\emph{Are data-driven power system control and operation methods robust against adversarial attacks?}

Unfortunately, in this paper, we show that in several benchmark power system control tasks, a minimal information attacker (e.g., attacker without any model information about the grid and the operation algorithm)
leads to unstable operation and deteriorates ML agent performance.
%In this paper, standing from the role of an adversarial attacker, we answer the question with a negative view for many currently proposed data-driven controllers. With minimal assumption on the attacker's access to system and ML models, 
In particular, we design a black-box, hard-to-detect data perturbation attack over the system measurements through state estimation or communication networks. On one of the test case of voltage regulation, we show that by only perturbing the user demand profiles and renewable generation values within $5\%$, the reactive power injection output from the trained RL agents can deviate a lot from the original RL decisions, thus the subsequent power flow profiles go beyond the safe operation range by more than 3 times compared to model-based counterparts.

This work is the first attempt to formalize the overlooked security threats brought by the integration of machine learning (or so-called data-driven) control methods into power grid operations. 
%We focus on evaluating cyberspace data perturbation attacks on learning agents along with \emph{undesired physical grid impacts}. 
We focus on evaluating the performance of RL agent against perturbations on its state observations, in terms of both action deviation and the physical system consequences it leads to. We look into RL agents trained 
on different power grid datasets (e.g., power demand, battery state-of-charge, power grid topology) and past decision sequences (e.g., power dispatch, battery charging/discharging signals, line switching records). 
Most RL agents find a near optimal action sequence,
either based on value function methods or policy gradient updates~\cite{schulman2017proximal, mnih2015human}, 
though it is challenging to interpret the learnt policy. %draw physical interpretations from such decision process. 
The attacker then make use of historical measurements and decision data from the RL-agent to manipulate system states adversarially. 
We also release the developed simulation testbed and attack examples for future research on machine learning security for power grid operations\footnote{\url{https://github.com/chennnnnyize/PowerSystem_RL_Attacks}}. This work calls for identification of security threats in developing data-driven control schemes.
 
%However, in order to achieve safe and reliable integration of such data-intensive algorithms, there is an urgent need to validate whether the use of both machine learning models and datasets are robust against malicious behaviors such as data injection attacks and poisoning attacks.  Such observations pose questions on the applicability of such learning-based agents into power systems, as the safety requirements of critical infrastructures may not be satisfied. In this paper, 

\subsection{Related Work}
The idea of exploiting machine learning for power system operation and control dates back to 1980s, where rule-based algorithms are learned to design autonomous decision makers~\cite{wollenberg1987artificial}. Recent advancements in RL algorithms hold the promises of scaling up to more realistic grid control settings. In addition, compared to rule-based algorithms, RL methods can learn directly from experiences without constructing an expert dataset~\cite{chen2021reinforcement, ernst2004power}. Learning-based control paradigms have been proposed for a variety of operation tasks in power grids, including control of voltage and frequency~\cite{xu2019optimal, daneshfar2010load, chen2020data, cui2020reinforcement}, 
capacity scheduling of PV and energy storage~\cite{huang2020deep}, topology control~\cite{marot2020learning} and many more. A more detailed review of RL for power system operation can be found in~\cite{chen2021reinforcement}. 

Power system research have discussed grid security issues in state estimation and physical threats such as communication failure and adverse weather conditions~\cite{morison2004power}, while few research studied the security issues brought from the emerging data pipelines and ML models. It is well worth noting that compared to recent research in robust RL~\cite{pinto2017robust}, we focus on the physical consequences of ML models, while in \cite{omnes2021adversarial}, a random attack is considered for reinforcement learning agent, while \cite{zheng2021vulnerability} examines the vulnerability of discrete actions learned by Deep Q-Network~(DQN) on a specialized topology control task, and the potential threats are valid when the weights of DQN are fully exploited. Our proposed attack framework can be generalized to a variety of learning-based controller settings, and the black-box formulation demonstrates such security breach is of real existence.

%% file: system.tex
In this section, we introduce and formulate a set of operation tasks in power systems in a general form. We will illustrate how such control tasks can be fitted into a learning framework with measurement data in Section~\ref{sec:RL}.

In many power system control tasks, the system operator seeks to minimize the operation cost for a certain period of time, subject to the system dynamics. Without loss of generality, we use $\bm{s}(t)$ to denote the observable system state at time $t$, such as line flow, voltage magnitude and frequency; $\bm{a}(t)$ to denote the collection of control actions which the system operator takes at time $t$, such as real and reactive power injections; and $\bm{v}_{ext}(t)$ to denote external/environment inputs such as renewable generation, time-varying demand. Throughput the paper, we use $\bm{s}(t), \bm{a}(t)$ and $\bm{s}, \bm{a}$ interchangeably. 
%in a vector $\bm{v}_{ext}(t)$, and 
Let $f(\cdot,\cdot,\cdot)$ represent the system time evolution,
\begin{equation}
    \boldsymbol{s}(t+1)=f\left(\boldsymbol{s}(t), \boldsymbol{a}(t), \bm{v}_{ext}(t) \right)
    \label{equ:state}
\end{equation}
%where we use $\bm{a}(t)$ to denote the collection of control actions which system operators take at time $t$, and $\bm{s}(t)$ to denote the observable system states, such as line flow, voltage magnitude and frequency. 
In order to find a sequence of optimal action sequence $\{ \bm{a}^*(0),\bm{a}^*(1),...\bm{a}^*(T)\}$ that minimizes the operation cost, we can formulate and solve the optimization problem,
\begin{subequations}
\label{equ:operation}
\begin{align}
  \min_{\bm{a}(t)} & \quad 
   \sum_{t=1}^T C(\bm{s}(t),\bm{a}(t))\\
  s.t. \; & \quad \text{system dynamics}~\eqref{equ:state} \\
  & \quad g(\bm{s}(t),\bm{a}(t), \bm{v}_{ext}(t))=\bm{0}; \label{equ:g}\\
  & \quad h(\bm{s}(t),\bm{a}(t), \bm{v}_{ext}(t))\leq \bm{0}; \label{equ:h}
\end{align}    
\end{subequations}
where cost function $C(\cdot, \cdot)$ is defined on both system states and control efforts; Constraint \eqref{equ:g} collects equality conditions such as nodal power balance and power flow equations; Constraint \eqref{equ:h} encodes hard system operating conditions, such as line flow limits and nodal voltage/frequency constraints.

In practice, we can use optimization solvers or heuristics to find a good solution for Problem \eqref{equ:operation}. But the lack of accurate physical model $f(\cdot), g(\cdot), h(\cdot)$ limits the applicability of such optimization framework. Moreover, the fast-changing renewable generation and more complex and distributed nature of the power grids bring computational challenges to solve large-scale problems in a fast timescale.

%% file: rl.tex
Faced with the increasing system complexity and the need for efficiently finding accurate solutions for a wide variety of decision-making tasks, data-driven controllers become a good fit.
Reinforcement learning provides a promising tool to address the aforementioned challenges. Specially, RL methods do not need knowledge of explicit power system models and can learn from interactions with the underlying system. In addition, once a policy is learnt, it does not require extra computation during execution.
In this section, we start by providing a brief review on the RL formulation, and then illustrate the potential issue of RL against adversarial attacks.

In the RL framework, we define the underlying power system models as a Markov Decision Process (MDP), which can be represented as a 5-tupe $(\mathcal{S}, \mathcal{A}, R, p, \gamma)$. At time step $t$, agent receives the state vector $\bm{s}(t) \in \mathbb{R}^{N}$, which is composed of various measurements from the power grids, and take action $\bm{a}(t) \in \mathbb{R}^{M}$ such as reactive/active power set-points of inverters, topology switch and etc. The environment, in return, takes the state-action pair $(\bm{s}(t), \bm{a}(t))$ as inputs and outputs the next state $\bm{s}_{t+1}$ based on the transition probability $p: \mathcal{S} \times \mathcal{A} \rightarrow \mathcal{P}(\mathcal{S})$, and a reward signal $r$ is given by reward function $R: \mathcal{S} \times \mathcal{A} \times \mathcal{S} \rightarrow \mathbb{R}$. %($\mathcal{P}(\mathcal{S})$ defines the set of all possible probability measures over state space)

The goal of a RL agent is to find a policy $\bm{\pi}(\bm{s}|\bm{a})$ to maximize the cumulative reward. In the value-based RL framework such as Q-learning~\cite[Chapter 6]{sutton2018reinforcement}, we fit a parameterized function $Q(\bm{s},\bm{a})$ which reflects the accumulated reward:
\begin{equation}
    Q_{\bm{\pi} }(\bm{s}, \bm{a})=\mathbb{E}_{\pi}\left[\sum_{k=0}^{\infty} \gamma^{k} r(t+k+1) \mid \bm{s}(t)=\bm{s}, \bm{a}(t)=\bm{a}\right].
\end{equation}
where $\gamma \in [0,1]$ is the discount factor. Let $Q^*(\cdot,\cdot)$ denote the optimal action-value function, then the optimal deterministic policy is given by 
\begin{equation}
\label{eq:value_based}
    \pi^{*}(s)=\arg \max_{a \in \mathcal{A}} Q^{*}(\bm{s}, \bm{a}).
\end{equation}

While in the policy-based RL framework~\cite[Chapter 13]{sutton2018reinforcement}, we directly train a policy neural network $\pi_\theta$ that maps the state to optimal action,
\begin{equation}
\label{equ:policy_based}
    \theta^{*} \in \arg \max _{\theta \in \Theta} J(\theta); \; J(\theta)=\mathbb{E}\left[\sum_{t=1}^{T} r\left(\bm{s}(t), \bm{a}(t)\right)\right],
\end{equation}
where $J(\theta)$ is the expected return of the current policy. We can use neural networks to fit the $Q$ function in the value-based RL framework, which leads to the development of Deep Q Network~(DQN)~\cite{mnih2015human}. In addition, recent works such as 
%Deep Deterministic Policy Gradient (DDPG)~\cite{lillicrap2015continuous}, 
Proximal Policy Optimization (PPO)~\cite{schulman2017proximal} and Advantage Actor Critic (A2C)~\cite{mnih2016asynchronous} learns both a critic network for estimating the expected reward in Eq~\eqref{equ:policy_based} and an actor network $\pi_\theta(\bm{a}|\bm{s})$ together, which lead to state-of-the-art performance.
%while in policy-based RL algorithms such as  and Advantage Actor Critic~(A2C, which also utilizes a critic neural network to learn Q value)~\cite{mnih2016asynchronous, schulman2017proximal}, a neural network $\pi_\theta$ can be also trained based on Equ. \eqref{equ:policy_based}.

RL algorithms provide an efficient alternative to the traditional model-based control/optimization approach, where it learns through interacting with the environment rather than knowing the system model. It can be applied to many control tasks of different timescales, by a proper choice of the reward function to reflect the operation cost, and penalize infeasible actions, power losses or deviations from the desired states. Note that in the standard RL framework, physical constraints such as line flow limits, voltage magnitude constraints are not modeled explicitly, while the agent is trained to explore in the environment to achieve high rewards while avoiding infeasible control actions. Since physical constraints in power grids are not directly incorporated into RL training schemes, this motivates us to examine if normally trained RL controllers will be compromised under adversarial observations.

%when only measurements but not the whole configurations/grid parameters are available. 

%Note that there is a distinction between the models defined in the RL framework, i.e., Eq.~\eqref{eq:value_based} or Eq.~\eqref{equ:policy_based}, and the models used in the real power grid. 

%% file: attack.tex
In this section, we start from the modeling assumptions of the attackers, and show a possible attack scheme even though the learning-based controllers' parameters are never compromised, while resulting state perturbations may lead to malicious operation decisions.

\subsection{Attacker's Knowledge and Capabilities}
\label{sec:attack_knowledge}
From an attacker's perspective, since deep RL agents have complex model architectures and model weights, and system operators often keep a secure copy of the control algorithm, it is thus hard to directly compromise the RL model. The ``black-box" nature of the control systems of many elements of the power system is well known (photovoltaic smart inverters, for example).  While emerging standards set broad guidelines for control system performance, the precise definition of the internal control loops in many systems are often traded as trade secrets.  Thus, it is reasonable to assume that cyber attackers may not have the ability to directly compromise the RL model and would resort to false data injection attacks on controller inputs to execute their attacks. 
To mount successful black-box data injection attacks, we present a novel mechanism that treats the unknown control map $\pi(\bm{s})$ as a differentiable function, and we assume the attacker can observe the agent's state observations. We regard such setting as a realistic one, since the attacker only needs \emph{query} access to the RL agent and can observe publicly available grid states. 

The resulting attack action is a small perturbation with respect to $\bm{s}$, such that it can bypass the bad data detection scheme and can mount to the RL agent's inputs. At timestep $t$ and given clean system measurements $\bm{s}_t$, we aim to construct a perturbed adversarial input $\bm{s}_{adv}(t):=\bm{s}(t)+\delta(t)$, such that the system performance is not optimal when the RL agent output actions $\bm{\pi}(\bm{s}_{adv}(t))$.

\subsection{Attack Implementation}
Essentially, the attacker tries to act against the power grid's operation goal based on his/her observation of the system states. From the viewpoint of machine learning algorithm, this is \emph{opposite} to the training objective, which the attacker is interested to find state variables rather than action variables that can minimize the reward. In the following, we will firstly formulate two forms of attacks based on the attacker's knowledge and capacity assumption described in Section \ref{sec:attack_knowledge}, and then describe how to find attack vectors by an approximated solution algorithm.

\textbf{Action Distortion Attack}\quad
In this setting, the attacker tries to distort the output actions $\bm{a}$ as much as possible, by injecting bounded state perturbation $\bm{\delta} (t)$,
%so that the system's operation scheme is deviating from the optimal one:
\begin{subequations}
\label{equ:distortion}
    \begin{align}
\max_{\bm{\delta} (t)} \quad & d( \pi\left(\bm{s}(t)+\bm{\delta}(t)\right),\pi\left(\bm{s}(t)\right)) \\
s.t. \quad & \norm{\bm{\delta} (t) }_p \leq \epsilon, \label{equ:norm}
\end{align}
\end{subequations}
where $d(\cdot, \cdot)$ is a distance measure between the original policy output and the policy output under adversarial attack. In the case of discrete action output, we define $d(\cdot, \cdot)$ as the cross entropy loss between the adversarial probability distribution and optimal policy generated by the RL agent. In the case of continuous action output, we define the distance as the $L2$ norm between the output policies. Constraint \eqref{equ:norm} ensures the injected measurement perturbation is small enough so that it can remain undetected. When $p=0$, such constraint describes number of compromised state entries.

\begin{algorithm}[]
\caption{Black-Box Attacks on RL Agents}
  \label{algo}
  \KwInput{Clean measurements $\bm{s}_0,...,\bm{s}_T$, pre-trained DRL policy $\bm{\pi}$}
  \KwOutput{Adversarial state perturbations $\delta_0,..., \delta_T$}
  \KwData{Maximal perturbation $\epsilon$; Attack goal $L$ (distorting actions or manipulating states)}
  \For{t=0,..., T}{
  $\bm{a}_0(t)=\bm{\pi}(\bm{s}(t)); \; \bm{s}_{adv}(t) \leftarrow \bm{s}(t).\texttt{copy}()$; \\
   \While{$\norm{\bm{s}(t)-\bm{s}_{adv}(t)}\leq \epsilon$}
   {  
     \For{i=0,..., N}{
   		Find attack direction $\hat{g}_i$ via Equation \eqref{equ:graident};\\
   		}
   		Perturb state space using Equation \eqref{equ:perturbation};\\
    Optional: Evaluate controller output $\bm{a}_{adv}(t)$ using $\bm{s}_{adv}(t)\leftarrow \bm{s}^{(j)}$;\\
   }
   $\bm{a}_{adv}(t) \leftarrow \pi\left(\bm{s}_{adv}(t)\right)$\;
$\bm{s}(t+1) \leftarrow \text {\texttt{env}.step }\left( \bm{a}_{adv}(t)\right)$
   }
\end{algorithm}

\textbf{Grid Manipulation Attack}
Under this setting, the attacker aims to manipulate the system state trajectory towards certain target state trajectory $\{\bm{s}_{target}(t)\}_{t=1}^{T}$. For instance, in the building energy management task, the attacker may want to maliciously take control to increase/decrease a certain room's temperature to an uncomfortable level as $\bm{s}_{target}(t)$; or in the frequency regulation task, the adversary may want to drive certain node's frequency towards a unsafe value as $\bm{s}_{target}(t)$. By considering system dynamics $f(\bm{s}(t), \bm{a}(t))$, we can formulate such manipulation attack as follows:
\begin{subequations}
\label{equ:manipulation}
\begin{align}
\min_{\bm{\delta}(1)...\bm{\delta}(T)}  \quad & \sum_{t=1}^{T} d\left(\bm{s}(t), \bm{s}_{target}(t)\right) \\
\text { s.t. } \quad & \bm{a}(t)=\bm{\pi}\left(\bm{s}(t)+\bm{\delta} (t)\right),\\ &\bm{s}(t+1)=f\left(\bm{s}(t), \bm{a}(t)\right),\\
\quad & \norm{\bm{\delta} (t) }_p \leq \epsilon, \forall t \in [1, ..., T].
\end{align} 
\end{subequations}
%\todo{Yize: should the optimization variable be a sequence of  $\delta(t)$? i.e., $\delta(1)...\delta(T)$? or you mean to inject the same noise to the entire trajectory?}

However, both Problem \eqref{equ:distortion} and \eqref{equ:manipulation} require solving an optimization problem which are highly nonconvex due to the existence of (deep) neural network policy $\bm{\pi}(\cdot)$. Further, the attacker does not have the exact model of system dynamics $f$ and system operation policy $\bm{\pi}(\cdot)$. Thus, it is challenging to find the exact solutions for these two problems. Recent work~\cite{goodfellow2014explaining} found that neural networks are highly sensitive to the adversarial perturbations of inputs. By taking only a few gradient steps with respect to the adversarial objective, the output of deep learning models will change a lot. This motivates us to find an approximate solution of \eqref{equ:distortion} and \eqref{equ:manipulation} via gradient method, and examine if such approach can achieve control action distortion or grid manipulation.

In our proposed scheme, we make use of the query access to approximate the objective's gradient with respect to the black-box RL model. For the $i$-th dimension of the state, we can approximate the gradient of the adversary's goal $L(\cdot)$ with respect to the input features\footnote{The choice of $L$ function depends on the attacker's goal and constraints, and we transform the constrained optimization problem \eqref{equ:distortion} and \eqref{equ:manipulation} into unconstrained form along with projection step.},
\begin{equation}
\label{equ:graident}
    \hat{g}_{i}:=\frac{\partial L(\mathbf{s})}{\partial s_{i}} \approx \frac{L\left(\mathbf{s}+h \mathbf{e}_{i}\right)-L\left(\mathbf{s}-h \mathbf{e}_{i}\right)}{2 h},
\end{equation}
where $\mathbf{e}_i$ is 1 only in the $i$-th entry and all 0s everywhere else. $h$ is a small constant to control the gradient step.

To find $\bm{s}_{adv}$, we iteratively take gradient steps to optimize the objective for the attacker. In practice, at iteration $j$, we also need to constrain the attack vector to satisfy the norm constraints. With $\Pi$ denoting the projection operator back to the norm constraints, we find the attack vector with the approximated gradient \eqref{equ:graident} by doing projected gradient descent,
\begin{equation}
\label{equ:perturbation}
    \bm{s}^{(j)}=\Pi_{\left[\bm{s}-\epsilon, \bm{s}+\epsilon\right]}\left(\bm{s}^{(j-1)}-\eta \cdot \operatorname{sign}\left(\hat{\bm{g}}^{(j)}\right)\right).
\end{equation}

We want to point out that the accuracy of the gradient estimation step will not be a major factor in the attack's performance as long as the sign (gradient direction) is correct. This has been showed in literature such as fast gradient sign method~(FGSM)~\cite{goodfellow2014explaining}, which only requires the sign (rather than the exact value) of the gradient to find adversarial examples, 

Note that in the proposed attack scheme, the direct access to the physical states, e.g., altering nodal voltage or cutting off power line is not allowed. The environment still transits from the true state $\bm{s}(t)$ rather than $\bm{s}_{adv}(t)$ to the next state $\bm{s}(t+1)$. To solve \eqref{equ:manipulation}, the attacker also needs the dynamics model $f(\cdot, \cdot)$. In this work, we focus on the case of one-shot grid manipulation attack, and attack one specific line/bus state, where we do not need the exact state transition model. In \cite{lin2017tactics}, further discussion on the timing of attacks is made, while more systematic modeling of the state model $\bm{s}_{t+1}=f((\bm{s}(t), \bm{a}(t))$ is discussed in \cite{weng2019toward}, which can craft stronger attack when the environment model can be learned by the attacker.

\subsection{On the Practicality of Attacks}
We would note that the proposed security threats are real concerns.
It has been reported that the sensing and communication networks (e.g., PMU and SCADA
network) are prone to attacks and can be compromised. For example, in the cyber-attack demonstrated by Idaho National Laboratory~\cite{Idaho}, ransomware demand could be made credible by manipulating protective relays to cause generator losing synchronization. 

The proposed attack does not interfere with the standard training process (e.g., directly modifying model weights of the RL agent). For real-world RL applications, the adversary can also be viewed as the worst case measurement noise or state estimation uncertainty. It is thus important to exam the robustness of RL agents under adversarial attack settings, before deploying it in safety-critical systems like power systems.  %and to include system dynamics and constraints in evaluating RL performance.

\begin{figure*}[t]
	\centering
	\includegraphics[width=0.95\linewidth]{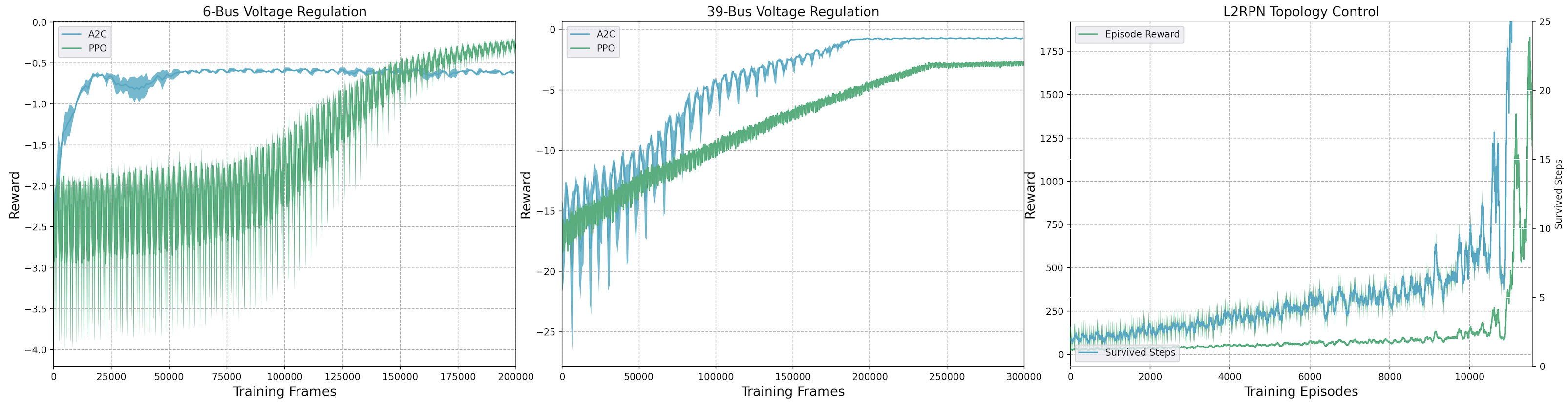}
	\caption{\footnotesize The training curve for DRL agents on three power system operation tasks.}
	\label{fig:training}
\end{figure*}

%% file: result.tex
In this section, we test a set of learning-based controllers on power system operations tasks under adversarial settings. Throughout the simulations, we show: 
\begin{itemize}
    \item Attackers with limited information about the learning algorithms can easily construct adversarial perturbations;
    \item Successful adversarial attacks have severe impacts, which may lead to infeasible operating conditions, loss of equipment, and large-scale load-shedding.
\end{itemize}  

All the DRL algorithms are implemented with PyTorch and the simulation environments are interfaced with OpenAI Gym, which are all open sourced in our Github repository. These benchmark tasks could provide standard comparisons on the security of learning-based power system controllers.

\subsection{Learning Task Description}

\newcommand{\bftab}{\fontseries{b}\selectfont}
\begin{table*}[h]
	\renewcommand{\arraystretch}{1.4}
	\centering
	\begin{tabular}{>{\centering\arraybackslash}m{3em}|>{\centering\arraybackslash}m{6em}|>{\centering\arraybackslash}m{6em}|>{\centering\arraybackslash}m{6em}|>{\centering\arraybackslash}m{6.5em}|>{\centering\arraybackslash}m{6em}|>{\centering\arraybackslash}m{6em}|>{\centering\arraybackslash}m{6em}|>{\centering\arraybackslash}m{6.5em}}
		\ChangeRT{1.0pt}
		Model    & \multicolumn{4}{c|}{6-Bus Example}& \multicolumn{4}{c}{IEEE 39-Bus} \\ 		\ChangeRT{0.7pt}
		
		& \multicolumn{3}{c|}{Average Reward}& {Distance} &\multicolumn{3}{c|}{Average Reward} & {Distance}\\
		\ChangeRT{0.7pt}
		%\hhline{~-------}
		Attack Type& No Attack & \texttt{Random} & \texttt{Action Distorted} & \texttt{State Manipulated} & No Attack & \texttt{Random} & \texttt{Action Distorted}&\texttt{State Manipulated}\\
				\ChangeRT{0.7pt}
		MPC &$-0.67\pm 0.9$ & \bm{$-0.71\pm 0.9$} & \bm $\bm{-0.73}\pm 0.9 $ & \bm{$-0.09\pm 0.08$} & \bm{$-1.41\pm 0.7$}& \bm{$-1.55\pm 0.8$}  & \bm{$-2.44\pm 0.9$} & \bm {$-0.07\pm 0.11$}\\
		PPO & \bm{$-0.13\pm 1.2$} & $-3.11\pm 4.2$ & $-5.91\pm 5.0$ & $-0.03 \pm 0.07$&$-2.91\pm 1.1$  & $-1.72\pm 1.8$ & $-15.44\pm 12.5$& $-0.04\pm 0.08$\\
		A2C & $-0.31\pm 0.6$ & $-0.76\pm 1.1$& $-3.06\pm 1.4$& $-0.03\pm 0.09$ & \bm{$-1.42\pm 1.0$} & $-2.22\pm 2.1$ &$-8.21\pm 6.1$ & $-0.03\pm 0.08$\\
		\ChangeRT{1.0pt}
	\end{tabular}
	\caption{Average rewards $\pm$ standard deviation for clean state no attack, random attack (\texttt{Random}), and action distortion attack (\texttt{Action Distorted}); and distance from targeted adversarial state (maximum active line flow limit)  $\pm$ standard deviation for state Manipulation attack (\texttt{State Manipulated}) over 200 test samples on the voltage regulation tasks. We choose $\epsilon=0.05$ with respect to normalized states.   }
	\label{table:voltage}
\end{table*}

\textbf{Voltage Regulation Task.} In this task, the RL agents are trained to optimize the  active and reactive power injections at buses with controllable generations and energy storage. Actions are continuous variables defined on power generators and inverters. We set up our power grid simulation based on Gym-ANM~\cite{henry2021gym}, and simulate on two test cases: a 6-bus toy example with synthetic load and renewables profiles, and an IEEE 39-bus test case with real-world data\cite{marino2016building}. The reward is a weighted sum of penalty on energy loss, overloaded lines and voltage violations. A large penalty is incurred when the AC power flow solutions diverge given the controller's setpoints. We trained two DRL agents, PPO~\cite{schulman2017proximal} and A2C~\cite{mnih2016asynchronous} to interact with both power grids.

\begin{figure}[t]
	\centering
	\includegraphics[width=0.8\linewidth]{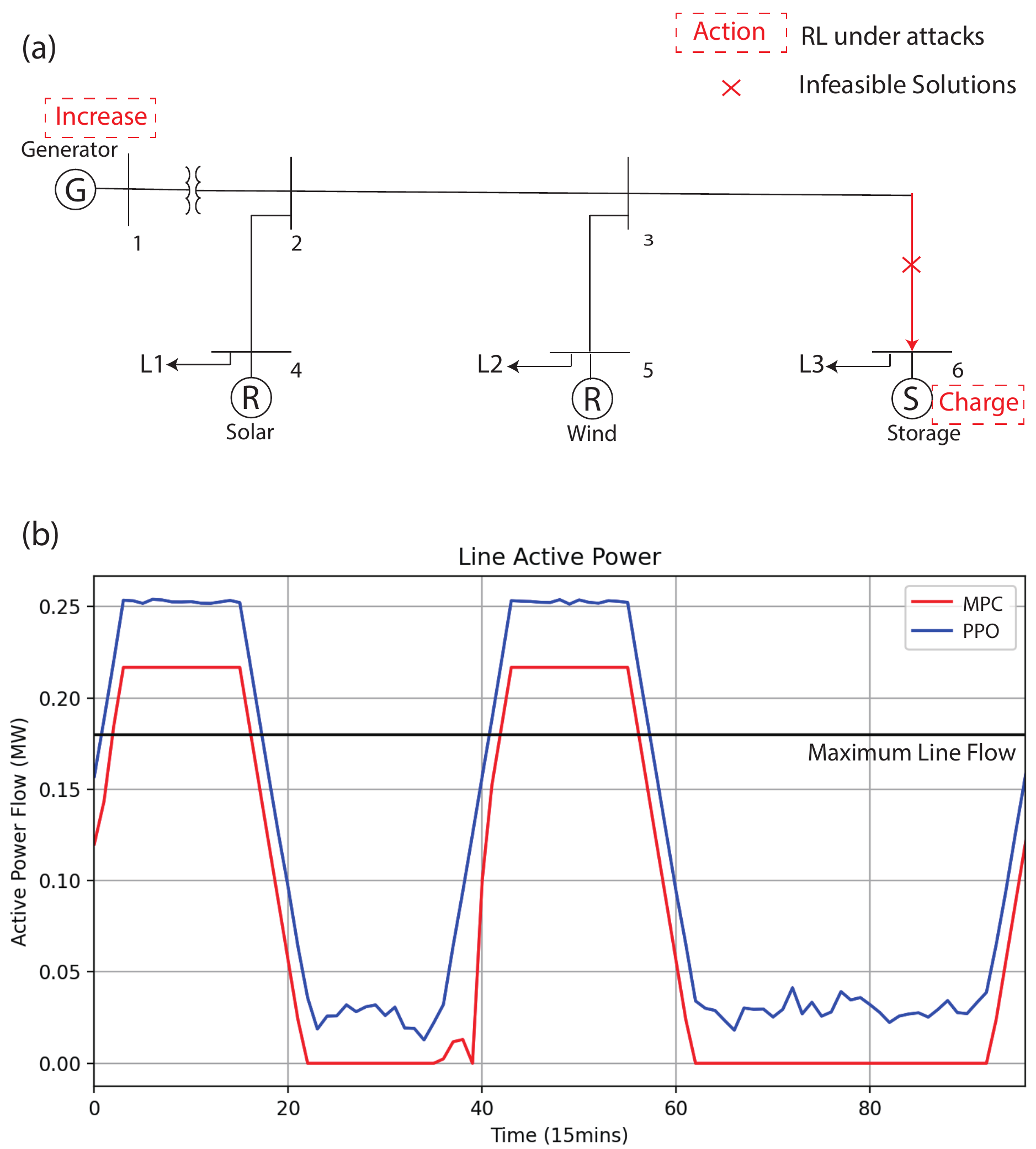}
	\caption{\footnotesize (a). The effects of \texttt{State Manipulated} attack, where the attacker targets on line $3-6$. The trained PPO agent is fooled by the adversarial state perturbations to take unsafe actions on energy storage and the generator, causing overloaded line. (b). Such adversarial attacks cause line overloading for more than 8 hours through 24 hours' test samples.}
	\label{fig:line}
\end{figure}

\textbf{L2RPN Topology Control Task.} We utilize the well-defined environment provided by the ``Learning to Run a Power Network'' competition~\cite{marot2020learning}, and investigate the topology reconfiguration task with actions on line switching, bus splitting and etc. The number of actions are exponential with respect to the number of lines and buses in the grid, making direct supervised learning impossible. For the details of features and discrete and continuous action set, we refer to the competition document~\cite{marot2020learning} and winning agent implementation~\cite{yoon2020winning}. The major objective of the L2RPN task is to operate the grid as many steps as possible, under time-varying demand and network topology. The reward function is defined by the available line margin, and a large penalty is enforced when cascading failures occur. Recent advances in DRL algorithms have enabled many agents that are specialized in achieving high reward in this competition~\cite{yoon2020winning}. 
While in this work, we investigate if such agents could still survive with slightly perturbed measurements coming from the environment. 
We trained a Double Dueling Deep Q Network~(D3QN)~\cite{van2016deep} to output discrete action set.

\textbf{Our Approach and Baselines} We set up our attack module by developing an intermediate layer between the power system environment simulator and RL algorithms. We only allow query access of the RL agent for the attack algorithm. While the attacker can sample clean system states from the environment, she is not allowed to inject any physical actions directly into the underlying grid. 
Future implementations can also use the intermediate layer to incorporate stronger adversaries, for instance, letting the adversary directly compromise the RL agents' weights, or training an RL adversary by interacting with the environment. The reward/survived steps evolution along with training steps are shown in Fig. \ref{fig:training}.

To benchmark the performance of the proposed attack methods, we set up a random attack scheme under perturbation norm constraints, where random noises with same $L2$ norm constraints are injected into state observations. For the voltage regulation task, we also compare with a model predictive controller (MPC) which has access to the DC approximation of the true system model, to validate the learned behaviors from DRL agents, and examine the vulnerabilities for such model-based counterparts. For the DRL training process, We kept a replay buffer $\left[\mathbf{s}_{t}, \mathbf{a}_{t}, r\left(\mathbf{s}_{t}, \mathbf{a}_{t}\right), \mathbf{s}_{t+1}\right]$ and ran the simulations on 3 random seeds. %since the topology is known while the future active power injections can be forecasted, 

\subsection{Simulation Results}
\subsubsection{Voltage Regulation}
We start by presenting the attack performance in the voltage regulation task, where we demonstrate the RL policies' deviations under different attack schemes.  
In Table. \ref{table:voltage}, we report the average reward for random attack and \texttt{Action Distorted} attack, as well as the distance from the targeted line power flow for the \texttt{State Manipulated} attack. 

In the clean testing environment (i.e., no attack), both the PPO agent and A2C agent achieve comparable performances to the MPC agent, which has a DC approximation of the true power flow model. All agents show some degree of robustness with respect to the $\texttt{Random}$ attack. However, the \texttt{Action Distorted} attack can pose severe threats to both RL agents. We observe the agents' actions can be far away from their actions on clean data, with only $5\%$ state perturbation in terms of $L2$ distance. More importantly, such impacted actions can be \emph{harmful} in both power system tasks, with frequent out-of-limit power flows and voltage magnitude observed. The trained RL agents can also be unstable, in the sense that the expected reward has much higher variance under attacks. It is thus important to characterize the decision boundaries made by RL faced with state uncertainties coming from renewable generation and sensor measurements. 

\begin{table}[h]
	\renewcommand{\arraystretch}{1.4}
	\begin{tabular}{>{\centering\arraybackslash}m{9em}|>{\centering\arraybackslash}m{5em}|>{\centering\arraybackslash}m{5em}|>{\centering\arraybackslash}m{6em}>{\centering\arraybackslash}m{6em}}
		\ChangeRT{1.0pt}
				& \multicolumn{3}{c}{L2RPN} \\
						\ChangeRT{0.7pt}
		Attack Type& No Attack & \texttt{Random} & \texttt{Action Distorted} \\
				\ChangeRT{0.7pt}
		Accumulated Reward & $202.38\pm 180.94$& $168.8\pm 366.0$&  42.14$\pm$ 56.27\\
    	Average Steps & $17.87\pm 137.7$ &$14.41\pm 495.1$ & 4.86$\pm$ 13.25  \\

		\ChangeRT{1.0pt}
	\end{tabular}
	\caption{Accumulated rewards $\pm$ standard deviation and average survival steps $\pm$ standard deviation for clean state no attack, random attack (\texttt{Random}), and action distortion attack (\texttt{Action Distorted}) over 200 test samples on the L2RPN topology control tasks.}
	\label{table:l2rpn}
\end{table}

\begin{figure}[t]
	\centering
	\includegraphics[width=0.9\linewidth]{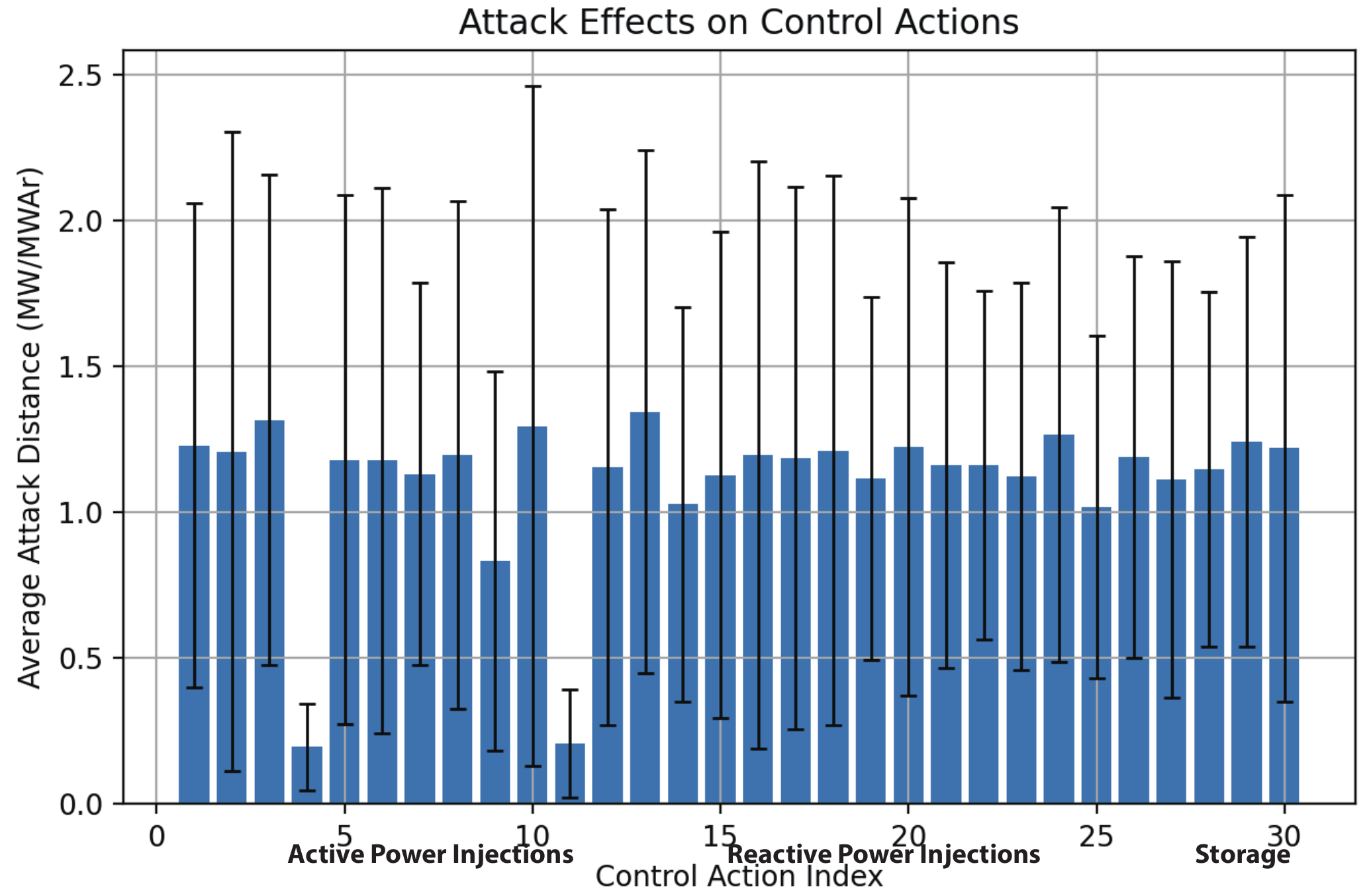}
	\caption{\footnotesize On the  effect of \texttt{Action Distorted} attack on the A2C algorithm for IEEE 39-bus grid voltage regulation task. Each bar denotes the average distance of actions taken under clean and adversarial states.}
	\label{fig:39bus}
\end{figure}

The results we report for \texttt{State Manipulated} attack in Table. \ref{table:voltage} are for targeted attack on a power line's active power flow, and the smaller value denotes greater distance from the line capacity. We can observe that on average MPC agent output actions can keep a larger margin from the safety boundary. Such attack impacts are further illustrated in Fig. \ref{fig:line}, which shows the line flow profile for a single line in the 6-bus simulation. The \texttt{State Manipulated} attack can cause storage at bus 6 to charge the battery during peak load period, causing overflow in line $3-6$.

\subsubsection{L2RPN Topology Control} The attack results for L2RPN task is reported in Table. \ref{table:l2rpn}. The data injection attacks can cause malicious changes on the $Q$ value outputs, and the actions under adversarial attacks cause frequent running failure in the grids, leading to over $75\%$ decrease in average reward and much shorter survival time. This illustrates that standard practice of RL agents may not be robust against data perturbations over high-dimensional measurements. We did not report \texttt{State Manipulated} results on L2RPN, as most of the time \texttt{Action Distorted} attack is detrimental and causing infeasible operation states.

\subsubsection{Discussions} 
\begin{figure}[t]
	\centering
	\includegraphics[width=0.99\linewidth]{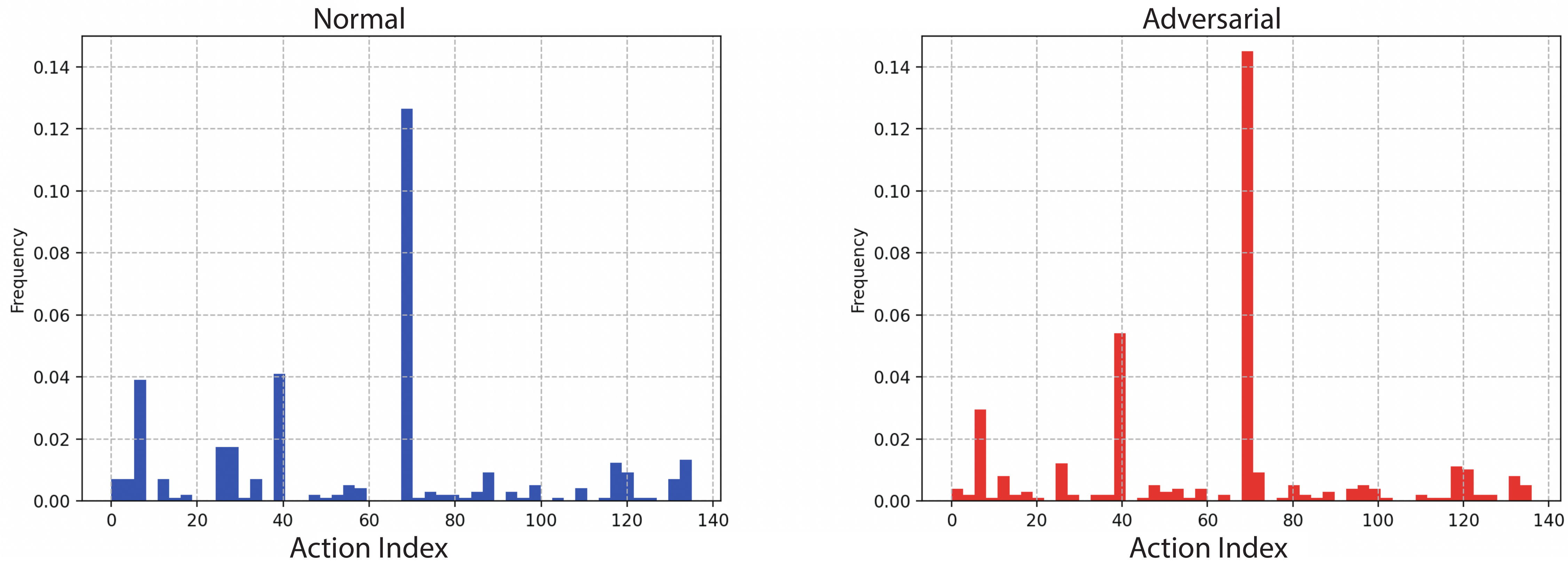}
	\caption{\footnotesize Empirical distribution of action choices for the L2RPN agent under clean power system observations (left) and under adversarial observation perturbation (right). }
	\label{fig:l2rpn}
\end{figure}

One interesting observation is that for both control tasks, some dimension of actions are more ``robust'' than others. For instance, we plot the statistics of RL agents' actions under \texttt{Action Distorted} attack in Fig. \ref{fig:39bus} and Fig. \ref{fig:l2rpn}, for voltage regulation tasks and topology control tasks respectively. In the IEEE 39-bus voltage regulation example, we find the active power injection on two generator buses are hardly impacted by attacks.
In the L2RPN tasks, the Q values for the top action sets are very similar, making it relatively easy to falsify the states to perturb action choices. %We hope the alternative analysis on the decisions made by learning-based controllers under attacks could 
Studying how to incorporate these observations to provide more design principles in robustifying nodes and lines which are easily to be attacked, or providing safety bounds against such input data uncertainties would be an interesting future direction.

We also want to comment on the computation complexity on finding the attack vectors. We observe the average time to implement Algorithm~\ref{algo} via query success for each step is within $0.5s$. While on the other hand, training DRL algorithms require over $100,000$ interactions with the environment and much more expensive computation.

%% file: conclusion.tex
In this paper, by designing and injecting adversarial measurement noises over system states with minimal assumption on attackers' knowledge and capabilities, we demonstrate that many learning-based closed-loop power system control laws are not robust. We show that in the power system community, research on ML approaches have mostly focused on their application practicability, while little attention have been made around safety issues. Compared with model-based techniques that rely on control or optimization methods which have more established notions on uncertainty and safety, we argue that including security objectives in learning-based controllers is essential to real-world power system tasks.  

\subsection{Identifying Practical Barriers on Implementing Learning-Based Controllers}
The ML community have taken great efforts on accelerating the training process of data-driven controllers, and state-of-the-art schemes have improved sampling performances and smoother learning  processes. Yet it is still not clear how the learned policies can be generalized to test environments, and how the decision boundary of data-driven controllers interact with physical constraints on power flow and devices. Moreover, current development of learning paradigms  heavily rely on synthetic test systems and public power system test cases, yet many implementation details and codes are not publicly available. Hence, to develop safe learning-based operational schemes, it is of urgent need to standardize benchmark datasets and learning algorithm testbeds including accurate and realistic grid dynamics. Sensitivities and safeguarding against false data injection attacks for ``black-box'' algorithms (from the perspective of the utility) should become part of the control comissioning process.  Emerging digital twin technology could also provide the means to monitor AI-based control systems post-comissioning to ensure reliable operation.

 %As AI-based control algorithms become more populous in power system control systems it will become even more critical to ensure that these systems are properly verified and comissioned prior to being brought online.  

Since reinforcement learning learns goal-oriented control laws from interaction with a system or a simulator, there is a necessity to develop, validate and investigate specific task environments, and achieve certain form of safety guarantees favorably in both agents' training exploration and real-world implementations.  And there is a strong incentive to develop computation efficient schemes that can reduce current need of tens of thousands of agents' interactive training steps while achieving fast transfer from simulated agents to real-world implementations.

\subsection{Security-Driven Algorithm Design}
There can be multiple design goals for a practical data-driven power system operator. And the agent is mostly trained in an ``optimistic'' fashion, in the sense that during training, grid failure is always allowed to explore unsafe operations with only penalty incurred; while during implementation, blackouts or constraint violations can not be provably avoided. Moreover, safety-critical processes have historically been largely static: what happened in one process cycle will happen exactly the same way in the next cycle.  Now, machine learning is being used to enable processes to be adaptive, to changes in the environment for example, the state after one run of the process can automatically influence the actions taken on subsequent runs often without a human in the loop. 
No longer is it merely the software's sources code or the physical grids that might contain vulnerabilities, but the machine learning model itself. Thus, vulnerabilities in hardware and software can still be present, but there is a new layer --- the model --- that is also now exposed and need to be investigated.

%For instance, in the L2RPN challenge, the training cost signal is given by a weighted sum of power loss, redispatch cost and grid failure indicator.  The recent developments on learning-based controllers have not yet considered the blackstart or grid restoration once severe damage occurs. 

%These processes are starting to be seen in many safety-critical places: the operation of the electrical power grid; transportation systems, including self-driving cars, aircraft avionics, and adaptive traffic control; manufacturing systems; medical sensors and health care delivery systems; and even large scientific facilities.

%Historically, static automated processes processes could be attacked in the same way that any hardware or software system could be attacked, by exploiting flaws in the design or bugs in the implementation that make the hardware or software vulnerable.  With machine learning and artificial intelligence, this process is becoming even more difficult.    That model can be attacked during the training phase, by ``poisoning'' the model with inputs that have undesirable effects, or during the operational phase when either through specially crafted inputs or brute force, an attacker similarly takes advantage of weaknesses in the model.   